# Ultrathin Ultra-broadband Electro-Absorption Modulator based on Few-layer Graphene based Anisotropic Metamaterial

Ayed Al Sayem[1]*, M.R.C.Mahdy[2]*, Ifat Jahangir[3], Md.Saifur Rahman[1]

[1]Dept. of EEE, Bangladesh University of Engineering and Technology, Dhaka, Bangladesh

[2] Department of Electrical and Computer Engineering, National University of Singapore, 4 Engineering Drive 3, Singapore

[3]Department of Electrical Engineering, University of South Carolina, Columbia, SC 29208, USA

[1]Dept. of EEE, Bangladesh University of Engineering and Technology, Dhaka, Bangladesh

*Corresponding Authors' Emails: A0107276@nus.edu.sg and ayedalsayem143@gmail.com

**Abstract: In this article, a few-layered graphene-dielectric multilayer (metamaterial) electro-optic modulator has been proposed in the mid and far infrared range that works on electro-absorption mechanism. Graphene, both mono layer and few layer, is an actively tunable optical material that allows control of inter-band and intra-band transition by tuning its chemical potential. Utilizing this unique feature of graphene, we propose a multilayer graphene dielectric stack where few layer graphene is preferred over mono layer graphene. Although the total thickness of the stack still remains in the nanometer range, this device can exhibit superior performances in terms of (i) high modulation depth, (ii) ultra-broadband performance, (iii) ultra-low insertion loss due to inherent metamaterial properties, (iv) nanoscale footprint, (v) polarization independence and (vi) capability of being integrated to a silicon waveguide. Interestingly, these superior performances, achievable by using few layer graphene with carefully designed metamaterial, may not be possible with mono layer graphene. Our proposals have been validated by both the effective medium theory and general transfer matrix method.**

## *I. Introduction*

High-speed electro-optic modulator is one of the key components in on-chip optical interconnects and optical communication systems [1]. These modulators are based on either electro refraction (ER) or electro absorption (EA) fundamentals, which are associated with the change in the real and imaginary parts of the refractive index respectively, under an applied voltage bias. In past, such modulators were designed exclusively with semiconductors, especially silicon [2, 3]. However, both ER and EA effects are weak in silicon at the telecommunication wavelengths and hence extremely large arm length is required for desired modulation depth. [4] The large footprint of semiconductor based optical modulator hinders its application in single chip integration. Sub-wavelength modulators are therefore highly desired in integrated photonic-electronic circuits. The key requirements to develop a modulator are nanometer scale footprint, low insertion loss, low switching energy, broadband performance, polarization independence and high modulation depth.

Graphene, a single atomic layer of carbon arranged in a hexagonal lattice [5], has drawn intense interest after being first discovered in 2004 [6-7]. Graphene has excellent mechanical strength, chemical stability, electro-optical tunability as well as the highest mobility of carriers (both electrons and holes) due to the unique conical band structure [5-7]. Although initially graphene was widely explored for applications in electronics, in recent years it has also shown promise as a key element in photonics [8], optoelectronics [9], plasmonics [10-11] and metamaterials [12-15]. Metamaterials [12-17] also have a deep impact on every recent research related with photonics. In both photonics and plasmonics, graphene has found frequent and extensive applications in electro-optic modulation [18-30]. In THz frequency range (0.1 to 10 THz), which is important for various applications [31-33], different types of graphene based electro-optic modulators have been proposed in recent years [18-22]. Graphene based electro-optic modulators have also been proposed at telecommunication wavelength [22-29] and in visible range [30].

In this theoretical study, we show that using few-layer graphene-dielectric multilayer anisotropic metamaterial, it is possible to obtain polarization independent high modulation depth (over 90%) with sub-wavelength size and ultra-low insertion loss for an ultra-broadband frequency range (from telecommunication band to high THz frequency). We also show why few layer graphene is more favorable instead of mono layer graphene in order to maximize graphene's potential in modulation applications. Another advantage of using few layer graphene is the ability to control the modulation depth by modifying the thickness of the modulator, which can be very crucial in many applications. Besides, this device is polarization independent and can be conveniently integrated into a silicon waveguide for modulation applications, unlike the case in [22, 27-29]. Finally, we briefly explain the physical phenomena governing the better performance achievable from this modulator using the anisotropic metamaterial theory, which is further verified by generalized $4 \times 4$ transfer matrix method (TMM) [34-36].

## II. *Theoretical Model*

Dependence of optical conductivity of monolayer graphene on chemical potential, temperature, frequency and relaxation time can be determined using the Kubo formalisms [37, 38] including both intra-band and inter-band contributions,

$$\sigma(\omega, \mu_c, \tau, T) = \frac{e^2}{\pi\hbar^2}\frac{i}{\omega - i\tau^{-1}}\left[\frac{1}{(\omega - i\tau^{-1})^2}\int_0^\infty \xi \left(\frac{\delta f_d(\xi)}{\delta\xi} - \frac{\delta f_d(-\xi)}{\delta\xi}\right)\delta\xi - \int_0^\infty \frac{f_d(-\xi) - f_d(\xi)}{(\omega - i\tau^{-1})^2 - 4(\xi\hbar^{-1})^2}\delta\xi] \ -\right] \quad (1)$$

The first and second parts of equation (1) correspond to the intra-band and inter-band contributions respectively. The intra-band contribution can be simplified as,

$$\sigma_{intra}\,(\omega, \mu_c, \tau, T) = \frac{-ie^2 K_B T}{\pi\hbar^2(\omega - i\tau^{-1})}\left[\frac{\mu_c}{k_B T} + 2\ln\left(e^{-\frac{\mu_c}{k_B T}} + 1\right)\right] \quad (2)$$

And the inter-band contribution can be approximated as ($\mu_c \gg k_B T$),

$$\sigma_{inter}\,(\omega, \mu_c, \tau, T) = \frac{-ie^2}{4\pi\hbar}\ln\left(\frac{2|\mu_c| - (\omega - i\tau^{-1})\hbar}{2|\mu_c| + (\omega - i\tau^{-1})\hbar}\right) \quad (3)$$

So the total conductivity of mono layer graphene can be given by,

$$\sigma_{mono}\,(\omega, \mu_c, \tau, T) = \sigma_{intra}\,(\omega, \mu_c, \tau, T) + \sigma_{inter}\,(\omega, \mu_c, \tau, T) \quad (4)$$

Where, $\omega$ is the angular frequency, $\mu_c$ is the chemical potential, $\tau$ is the relaxation time, e is the electron charge, $\hbar$ is the reduced plank's constant, $k_B$ is Boltzmann's constant, T is the temperature, $\xi$ is the energy and $f_d(\xi)$ is the Fermi Dirac distribution given by, $f_d(\xi) = \left(1 + e^{\frac{\xi - \mu_c}{K_B T}}\right)^{-1}$

One of the key findings of this work is the use of few layer graphene instead of monolayer graphene to drastically reduce the thickness of the modulator. It is already known that few layer graphene (N being the number of layers) has N-fold conductivity of monolayer graphene in THz, far infrared and visible range [29, 39, 40]. For multilayer graphene, low-energy electronic states of graphene are significantly controlled by interlayer interaction [41]. But they do not cause a significant change in the excitonic resonance energy [41]. For this reason in near infrared and in visible range, the in-plane optical conductivity of graphene layers increases almost linearly with the number of layers. So for few layer graphene($N < 10$), conductivity can be expressed as,

$$\sigma_{few}\,(\omega, \mu_c, \tau, T) = N\sigma_{mono}\,(\omega, \mu_c, \tau, T) \quad (5)$$

Because of its 2D nature, monolayer graphene is basically an optically uni-axial anisotropic material, whose permittivity tensor can be given by,

$$\varepsilon_{graphene} = \begin{bmatrix} \varepsilon_{g,t} & 0 & 0 \\ 0 & \varepsilon_{g,t} & 0 \\ 0 & 0 & \varepsilon_{g,\perp} \end{bmatrix} \quad (6)$$

Graphene's tangential permittivity can be given by,

$$\varepsilon_{g,t} = 1 + j\frac{\sigma_{mono}\,(\omega,\mu_c,\tau,T)}{\omega\,\varepsilon_o\,t_{gmono}} \quad (7)$$

where $\omega$ is the angular frequency, $\varepsilon_o$ is the free space permittivity and $t_{gmono}$ is the thickness of monolayer graphene. As graphene is a truly two dimensional material, the normal electric field cannot excite any current in the graphene sheet, so the normal component of the permittivity is given by $\varepsilon_{g,n} = 1$ [42]. However, for few layer graphene normal component of permittivity, $\varepsilon_{g,\perp}$ can be approximated as graphite's permittivity (~2), and tangential permittivity can be given by,

$$\varepsilon_{g_{fewlayer}\ ,t} = 1 + j\frac{\sigma_{few}\,(\omega,\mu_c,\tau,T)}{\omega\,\varepsilon_o N t_{gmono}} \quad (8)$$

Here the thickness of few layer graphene is N times the thickness of mono layer, $t_{gfew} = Nt_{gmono}$ . So from equation (5), (7) and (8) few layer graphene permittivity turns out to be the same as mono layer graphene [29, 43]. Although few layer graphene has the same permittivity as its mono layer counterpart, we still see performance enhancement in modulation, which will be explained below form anisotropic metamaterial approach.

At first we discuss the electro-absorption mechanism in graphene. There are two types of absorption mechanism in graphene: intra band absorption and inter band absorption. In this paper, we have concentrated on inter-band absorption within a frequency range from 180 THz to 482 THz (wavelength, $\lambda = 0.62\ \mu m\ to\ 1.67\ \mu m$ ). When the incident photon energy is greater than or equal to twice the Fermi energy $E_F$ of graphene, inter-band absorption occurs. By electrostatic doping, the Fermi energy $E_F$ can be tuned and the inter-band transitions with energies less than $2E_F$ can be blocked due to Pauli blocking [44].

Now, we obtain the effective permittivity tensor of (few/mono layer) graphene–dielectric multilayer metamaterial with a unit cell thickness of $d$ (one unit cell consists of few layer graphene and dielectric) which is between 5 nm and 25 nm. Within the specified frequency range, $k_o d \ll 1$, where $k_o$ is the free space wave vector. Therefore, from effective medium theory (EMT), we find the effective permittivity tensor of the multilayer metamaterial as,

$$\varepsilon_{eff} = \begin{bmatrix} \varepsilon_{||} & 0 & 0 \\ 0 & \varepsilon_{||} & 0 \\ 0 & 0 & \varepsilon_{\perp} \end{bmatrix} \tag{9}$$

$$\varepsilon_{||} = f\varepsilon_{g,t} + (1-f)\varepsilon_d \tag{10}$$

$$\varepsilon_{\perp} = \left(\frac{f}{\varepsilon_{g,\perp}} + \frac{1-f}{\varepsilon_d}\right) \tag{11}$$

Where $\varepsilon_d$ is the dielectric permittivity and $f$ is the fill fraction defined as,

$$f = \frac{t_g}{t_g + t_d} \tag{12}$$

Benefits of using few layer graphene can be easily understood from the concept of fill fraction. Although both few layer and mono layer graphene permittivity have same value, but the fill fraction for few layer graphene dielectric multilayer metamaterial is always higher than mono layer as per equation (12).

For a non-magnetic uni-axial metamaterial, propagating wave vector for s and p polarization can be given by,

$$k_{zs} = \sqrt{\varepsilon_{||}k_0^2 - k_x^2} \text{ (For s polarization)} \tag{13}$$

$$k_{zp} = \sqrt{\varepsilon_{||}k_0^2 - \frac{\varepsilon_{||}}{\varepsilon_{\perp}}k_x^2} \text{(For p polarization)} \tag{14}$$

Now, for an anisotropic metamaterial slab with thickness *d*, total magnetic fields for TM (p) polarized wave can be given by,

$$H_y^{incident\ \ media} = H_o e^{-(jk_z + jk_x)} + r_p e^{(jk_z - jk_x)} \tag{15a}$$

$$H_y^{Amm\ \ media} = H^+ e^{-(jk_{zp} + jk_x)} + H^- e^{(jk_{zp}z - jk_x)} \tag{15b}$$

$$H_y^{exit\ \ media} = t_p e^{-(jk_z + jk_x)} \tag{15c}$$

Where $k_o$ is the free space wave vector, $k_x = k_o sin\theta$ and $k_z = k_o cos\theta$, $\theta$ is the incident angle. $d$ is the slab thickness. $r_p$ and $t_p$ are the complex amplitude coefficients of reflection and transmission of TM (p) wave by the slab respectively. The four coefficients of equation (15) can be found by matching the boundary conditions for the tangential electric and magnetic fields at $z = 0$ and $z = d$ respectively

$$r_p = \frac{j\left(p-\frac{1}{p}\right)\sin\left(k_{2z}d\right)}{2\cos\left(k_{zp}d\right)+j\left(p+\frac{1}{p}\right)\sin\left(k_{zp}d\right)} \tag{16a}$$

$$t_p = \frac{2}{2\cos\left(k_{zp}d\right)+j\left(p+\frac{1}{p}\right)\sin\left(k_{zp}d\right)} \tag{16b}$$

Where $p = \frac{k_x \varepsilon_y}{k_{zp}}$. For a TE (s) polarized wave, we can easily obtain reflection and transmission coefficients similar to the case of TM polarized wave. For TE polarized wave, $p = \frac{k_x \mu_y}{k_{zs}}$.

From (16a) and (16b), power transmission and reflection coefficients can be obtained as,

$$T_P = \left|t_p\right|^2 \tag{17a}$$

$$R_P = \left|r_p\right|^2 \tag{17b}$$

Absorption coeffcient can be obtained as,

$$A_S = 1 - R_S - T_S \text{ (For s polarization)} \tag{18a}$$

$$A_P = 1 - R_P - T_P \text{ (For p polarization)} \tag{18b}$$

Since the definition of modulation depth has not been consistent in various reports in the literature [45], the following relations are the ones that we find suitable to calculate intensity (square of the amplitude of electric field) modulation using two different reference values, namely the minimum and maximum of the intensity ($T_{max}$ and $T_{min}$),

$$M_{Tmax} = \frac{T_{max} - T_{min}}{T_{max}} \times 100\% \tag{19a}$$

$$M_{Tmin} = \frac{T_{max} - T_{min}}{T_{min}} \times 100\% \tag{19b}$$

The first definition (equation (19a)) defines the modulation depth in the range of 0 to 100% and the latter definition (equation (19 (b)) in the range of 0 to infinity. We have adopted the first definition in this article because of its more intuitive nature [45].

Another important parameter for a modulator, insertion loss, can be defined as,

$$IL = 10\log_{10}(T_{max}) \tag{20}$$

## III. Discussion and Results

Although this device can act as a broadband modulator, we have generally concentrated on the telecommunication wavelength. But the same physics will also hold for a wide range of frequencies up to 482 THz which has also been shown later on.

Fig. 1 shows the schematic of the few layer graphene-dielectric multilayer metamaterial modulator (a) and its working mechanism (b). This modulator works on electro-absorption mechanism:

a) When $\mu_c < \frac{\hbar\omega}{2}$, inter-band transition (for both few layers and mono layer) is allowed and so absorption occurs. Depending on the number of unit cells and the loss per unit cell, total absorption can be controlled. This device, in all cases, has very low reflection because real part of tangential permittivity of the multilayer metamaterial is very close to ambient (air) permittivity. When the incident electromagnetic wave is totally attenuated in the device, the device is at off state as almost no transmission is allowed.

b) When $\mu_c > \frac{\hbar\omega}{2}$, inter-band transition is suppressed due to Pauli blocking [44] and intra-band absorption is only dominant in very low frequency (~1 THz) region. Therefore with the frequency range of interest, very high transmission is achievable for $\mu_c > \frac{\hbar\omega}{2}$ due to its sub-wavelength feature, negligible reflection and absorption. In fact, almost perfect transmission is achievable at on state which will be discussed below. As a consequence, insertion loss (key performance factor for any modulator) will be extremely low.

c) Highest chemical potential which have been achieved by electrostatic doping in graphene is $\mu_{cmax}$ =1 eV. So the upper bound of frequency can be found as $f = \frac{2\mu_{cmax}}{\hbar} = 482\ THz$

### A. Graphene's Conductivity

Fig. 2a shows mono and few layer graphene's conductivity (both real and imaginary parts) as a function of chemical potential at wavelength, $\lambda = 1.55\ \mu m$. Graphene's conductivity increases linearly with the number of layers. Inter-band contribution is dominant only for values of chemical potential lower than $\mu_c < \frac{\hbar\omega}{2}$. For $\mu_c > \frac{\hbar\omega}{2}$ Pauli blocking occurs and inter-band transition is suppressed. Fig. 2b shows few layer graphene's conductivity (N=3, both real and imaginary parts) for different chemical potential values as a function of frequency.

Fig. 2c shows graphene's tangential permittivity (both real and imaginary part) as a function of chemical potential at $\lambda = 1.55\mu m$. Fig. 2d shows few layer graphene's tangential permittivity (both real and imaginary parts) in the frequency range 180 to 482 THz for different chemical potential values. Inter-band contribution is dominant only for values of chemical potential $\mu_c < \frac{\hbar\omega}{2}$, which also makes the imaginary part of graphene's tangential permittivity dominantin that range. For $\mu_c > \frac{\hbar\omega}{2}$ imaginary part reduces drastically as intra band-transition dominates for $\mu_c > \frac{\hbar\omega}{2}$Loss in intra-band transition is dependent on graphene's scattering rate (inverse of relaxation time $\tau$) and is relatively very low.

Fig. 2e shows (few layer) graphene-dielectric multilayer anisotropic metamaterial's tangential permittivity as a function of chemical potential for different number of graphene layers in the unit cell, that consists of mono/few layer graphene and the dielectric. Although mono and few layer graphene's permittivity are same, fill fraction is higher for few layer graphene as described above. Again, similar to the case of (mono or few) graphene, the imaginary part of effective tangential permittivity of the anisotropic metamaterial also drastically changes at chemical potential $\mu_c = \frac{\hbar\omega}{2}$. But because of the higher fill fraction, for few-layer graphene, both real and imaginary part of effective tangential permittivity increase with the increasing number of graphene layers. Fig. 2f shows graphene-dielectric multilayer anisotropic metamaterial's tangential permittivity (both real and imaginary parts) in the frequency range from 180 to 482 THz for different chemical potential values.

### B. *Benefits of using few Layer graphene*

Benefits of using few layer graphene can be clearly observed from the EMT. Attenuation of electromagnetic wave inside the device can be calculated from the imaginary part of propagating wave vector $\Im(k_{zp})$ (equation (13) and (14)) in the (mono/few layer) graphene-dielectric multilayer structure. Fig. 3a shows $\Im(k_{zp})$ as a function of chemical potential for different number of graphene layers in a unit cell. Since few layer graphene-dielectric device has higher fill fraction as described before, $\Im(k_{zp})$drastically increases with the number of graphene layers in a unit cell, which determines the loss in the whole structure according to the EMT.

Another parameter playing an important role in reducing the modulator's thickness is the dielectric thickness in a unit cell. Fig. 3b shows $\Im(k_{zp})$ as a function of chemical potential for different dielectric thicknesses. Form effective medium theory (EMT), with reduced thickness of dielectric, fill fraction increases (equation (12)). As a result when $\mu_c < \frac{\hbar\omega}{2}$, the imaginary part of propagating wave vector increases with the decrease in dielectric thickness. This reduced dielectric thickness does not have any adverse effect in transmission whereas in [22], dielectric thickness played a critical role in transmission.

Fig. 3c shows $\Im(k_{zp})$ as a function of different dielectric thicknesses for two different values of chemical potential at telecommunication wavelength, $\lambda = 1.55\ \mu m$. For $\mu_{c1} = 0.35\ eV$, which is less then $\frac{\hbar\omega}{2} \approx 0.4\ eV$, inter-band absorption is dominant and as described above, $\Im(k_{zp})$ increases with reduced thickness. For $\mu_{c2} = 0.45\ eV$, which is greater then $\frac{\hbar\omega}{2} \approx 0.4\ eV$, inter-band absorption is suppressed and $\Im(k_{zp})$ is close to zero, irrespective of the dielectric thickness.

### C. *Modulation performance*

Fig. 4a shows the transmission, reflection and absorption characteristics of a nanomodulator (with L = 408 nm) as a function of chemical potential of graphene (other device parameters are given in figure caption). When $\mu_c < \frac{\hbar\omega}{2} \approx 0.4\ eV$, absorption is high and transmission is low. When $\mu_c > \frac{\hbar\omega}{2} \approx 0.4\ eV$, absorption is almost fully suppressed and transmission becomes high. At a certain chemical potential, maximum transmission is achieved. This can also be explained from the EMT. From equation (13) and (14) for normal incident, $k_{zs,p} = \varepsilon_{||} k_o$. So when $\varepsilon_{||} = 1$, no reflection should be observed. So, by increasing the chemical potential after crossing $\frac{\hbar\omega}{2}$, when the effective tangential (parallel) permittivity $\varepsilon_{||}$ equals to 1, there is a reflection dip resulting in high transmission. This high transmission is independent of the length of the modulator unlike [22]. As a result, insertion loss in this modulator is very low. Fig. 4b shows transmission (calculated using both $4 \times 4$ TMM and EMT) with effective parallel permittivity (equation (10)) and ambient media (air) permittivity, which validates our aforementioned theoretical analysis.

Fig. 5a shows the modulation depth calculated from equation (19a) (by both $4 \times 4$ TMM and EMT). Form the definition (19a), the highest modulation that can be achieved is 100%. From fig. 5a, we observe that more than 90% modulation depth can be achieved for an ultra-broadband frequency range with modulator size L=408 nm only. Fig. 5b shows the insertion loss (equation (20)) (by both $4 \times 4$ TMM and EMT). As described above, insertion loss in this modulator is extremely low.

Fig. 6a shows the modulation depth as a function of total thickness of the modulator. It can be observed that 60% modulation depth can be achieved with a 100nm thick modulator only (at telecommunication band, $\lambda = 1.55\ \mu m$). With the increase of the total thickness, modulation performance increases. Fig. 6b shows the insertion loss (equation (20)) as a function of total thickness of the modulator. As described above, insertion loss in almost negligible and independent of the modulator thickness.

## IV. *Numerical Method*

To take account into the anisotropic nature of graphene, we have used $4 \times 4$ transfer matrix algorithm [34-36] to calculate transmission, reflection and absorption of graphene based multilayer structure.

A general transfer matrix can be defined for any given layered structure [35-36]:

$$\begin{pmatrix} A_s \\ B_s \\ A_p \\ B_p \end{pmatrix} = T \begin{pmatrix} C_s \\ D_s \\ C_p \\ D_p \end{pmatrix} = \begin{pmatrix} T_{11} & T_{12} & T_{13} & T_{14} \\ T_{21} & T_{22} & T_{23} & T_{24} \\ T_{31} & T_{32} & T_{33} & T_{34} \\ T_{41} & T_{42} & T_{43} & T_{44} \end{pmatrix} \begin{pmatrix} C_s \\ 0 \\ C_p \\ 0 \end{pmatrix} \tag{21}$$

Where $A_p, A_s, B_p, B_s$ denote the complex amplitudes of p and s polarized waves of the incident and reflected waves respectively. $C_p$ and $C_s$ denote the complex amplitude of the transmitted waves in the exit medium. T is the same matrix used in [35] (equation 22). The general transfer matrix T can be defined as [36],

$$\boldsymbol{T} = \boldsymbol{L_a^{-1}} \prod_{i=1}^{N} \left[\boldsymbol{T_{ip}}(d_i)\right]^{\boldsymbol{-1}} \boldsymbol{L_f} = \boldsymbol{L_a^{-1}} \prod_{i=1}^{N} \boldsymbol{T_{ip}}(-d_i) \boldsymbol{L_f} \tag{22}$$

Where $\boldsymbol{T_{ip}}$ is the transfer matrix for ith layer and $\boldsymbol{L_a}$ is the incident matrix and $\boldsymbol{L_f}$ is the exit matrix. $\boldsymbol{T_{ip}}$ can be given by,

$$\boldsymbol{T_{ip}} \equiv e^{-\frac{j\omega}{c}\Delta d} \tag{23}$$

Where, *ω* is the angular frequency, *c* is the velocity of light in vacuum, *d* is the thickness of the i-th layer. For an non magnetic general anisotropic media whose permittivity tensor can be given by,

$$\boldsymbol{\varepsilon} = \begin{bmatrix} \varepsilon_{xx} & \varepsilon_{xy} & \varepsilon_{xz} \\ \varepsilon_{yx} & \varepsilon_{yy} & \varepsilon_{yz} \\ \varepsilon_{zx} & \varepsilon_{zy} & \varepsilon_{zz} \end{bmatrix} \tag{24}$$

The Δ matrix in equation (22) can be calculated as [36],

$$\Delta = \begin{pmatrix} -k_x \frac{\varepsilon_{zx}}{\varepsilon_{zz}} & -k_x \frac{\varepsilon_{zy}}{\varepsilon_{zz}} & 0 & 1-\frac{k_x^2}{\varepsilon_{zz}} \\ 0 & 0 & -1 & 0 \\ \varepsilon_{yz}\frac{\varepsilon_{zx}}{\varepsilon_{zz}} - \varepsilon_{yx} & k_x^2 - \varepsilon_{yy} + \varepsilon_{yz}\frac{\varepsilon_{zy}}{\varepsilon_{zz}} & 0 & k_x \frac{\varepsilon_{yz}}{\varepsilon_{zz}} \\ \varepsilon_{xx} - \varepsilon_{xz}\frac{\varepsilon_{zx}}{\varepsilon_{zz}} & \varepsilon_{xy} - \varepsilon_{xz}\frac{\varepsilon_{zy}}{\varepsilon_{zz}} & 0 & -k_x \frac{\varepsilon_{xz}}{\varepsilon_{zz}} \end{pmatrix} \tag{25}$$

Here $k_x$ is the tangential momentum of incident light and is given by,

$$k_x = n_a sin\varphi_{incident} \tag{26}$$

where $n_a$ is the refractive index of the incident media and $\varphi_{incident}$ is the incident angle.

The inverse of incident matrix and the exit matrix can be given by,

$$\boldsymbol{L_a}^{-1} = \frac{1}{2}\begin{pmatrix} 0 & 1 & \frac{-1}{n_a \cos\varphi_a} & 0 \\ 0 & 1 & \frac{1}{n_a \cos\varphi_a} & 0 \\ \frac{1}{\cos\varphi_a} & 0 & 0 & \frac{1}{n_a} \\ \frac{-1}{\cos\varphi_a} & 0 & 0 & \frac{1}{n_a} \end{pmatrix} \tag{27}$$

$$\boldsymbol{L_f} = \begin{pmatrix} 0 & 0 & \cos\varphi_f & 0 \\ 1 & 0 & 0 & 0 \\ -n_f \cos\varphi_f & 0 & 0 & 0 \\ 0 & 0 & n_f & 0 \end{pmatrix} \tag{28}$$

Where $n_f$ is the refractive index of the exit media and the angle $\varphi_f$ can be given by,

$$\cos\varphi_f = \sqrt{1-[(n_a / n_f)\sin\varphi_{incident}]^2} \tag{29}$$

The transmission and reflection amplitude coefficients of layered systems for both s and p polarized wave can be obtained in terms of the elements of the general transfer matrix T of equation (21).

$$t_{SS} \equiv \left(\frac{C_s}{A_s}\right)_{A_p=0} = \frac{T_{33}}{T_{11}T_{33}-T_{13}T_{31}} \tag{30a}$$

$$r_{SS} \equiv \left(\frac{B_s}{A_s}\right)_{A_p=0} = \frac{T_{21}T_{33}-T_{23}T_{31}}{T_{11}T_{33}-T_{13}T_{31}} \tag{30b}$$

$$t_{PP} \equiv \left(\frac{C_P}{A_P}\right)_{A_S=0} = \frac{T_{11}}{T_{11}T_{33}-T_{13}T_{31}} \tag{31a}$$

$$r_{PP} \equiv \left(\frac{B_P}{A_P}\right)_{A_S=0} = \frac{T_{11}T_{43}-T_{41}T_{13}}{T_{11}T_{33}-T_{13}T_{31}} \tag{31b}$$

Where $r_{SS}$ , $r_{pp}$ , $t_{SS}$ , $t_{pp}$ are the complex amplitude coefficients of the s and p polarized reflected and transmitted waves respectively. From which the power coefficients for the reflected and transmitted waves can be calculated as,

$$T_{ss} = |t_{SS}|^2 \tag{32a}$$

$$R_{ss} = |r_{SS}|^2 \tag{32b}$$

$$T_{pp} = \left|t_{pp}\right|^2 \tag{33a}$$

$$R_{pp} = \left|r_{pp}\right|^2 \tag{33b}$$

Finally absorption coefficients for s and p polarized waves can be given by,

$$A_s = 1 - T_{ss} - R_{ss} \tag{34a}$$

$$A_p = 1 - T_{pp} - R_{pp} \tag{34b}$$

## *V. Conclusion*

In this article, we have proposed a novel few layered graphene-dielectric multilayer (effective metamaterial) electro-optic modulator. Our theoretical investigation reveals that this device will have superior performance exhibiting almost every important criteria of a satisfactory modulator: extremely high modulation depth, ultra-low insertion loss, ultra broadband bandwidth, nano-scale footprint, polarization independence and the capability of being integrated to a silicon waveguide. It has also been established through our study that although the device exhibits superior performance in all the aspects of an electro-optic modulator, it still maintains a total size in the nanometer scale. The fundamental mechanism of this superior performance stems from the concept of using few layer graphene instead of mono-layer graphene with a careful metamaterial design, which has been validated using both effective medium theory and general transfer matrix method. We are very hopeful that this work would pave the way to the realization of a graphene based electro-optic modulator, performing at the highest limit achievable by this wonder material.

## Figures and Captions List

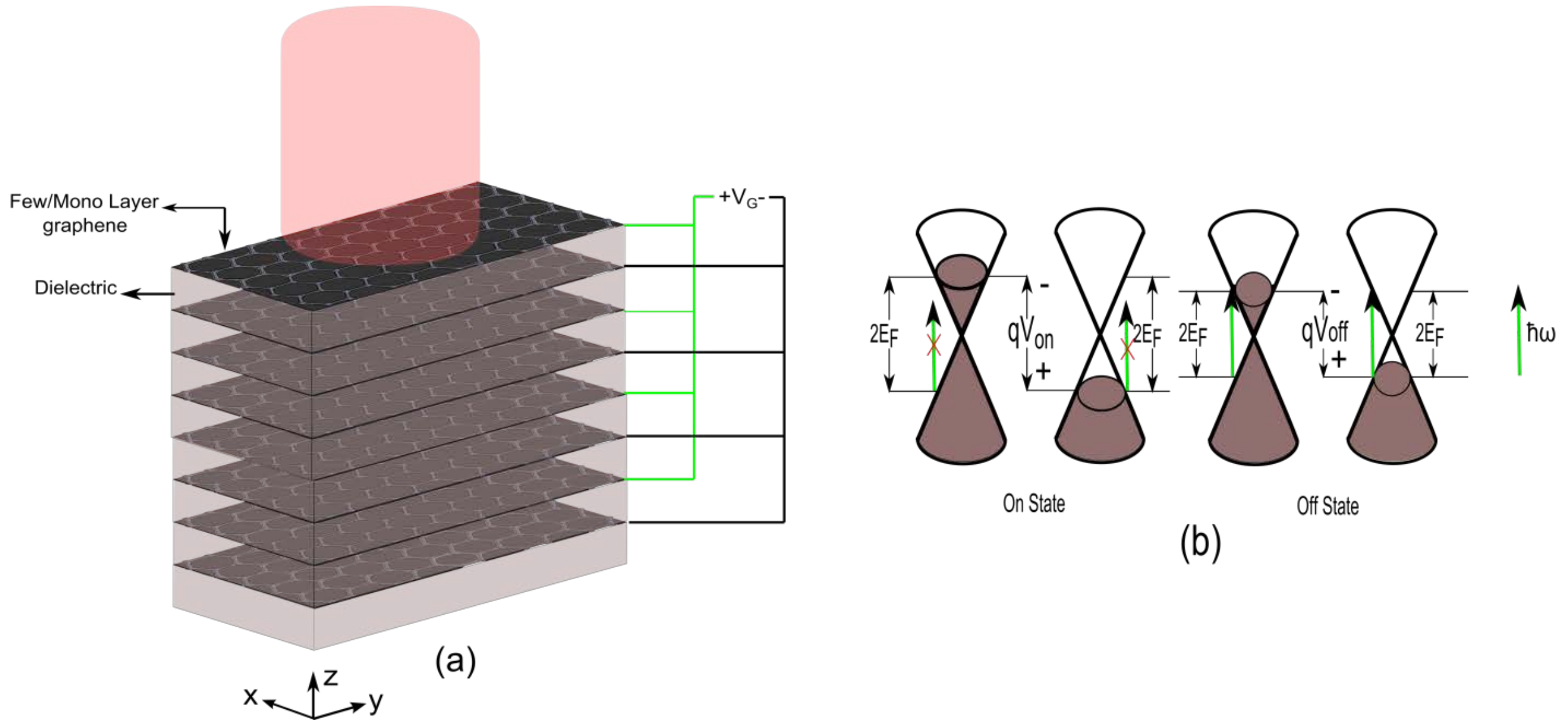

Fig. 1: (a) Schematic of the (few layer) graphene dielectric multilayer anisotropic metmaterial. (b) Band-diagram of the graphene layers at on and off state condition. At on state (high transmission) $\left(qV_{on} = 2E_f\right) > \hbar\omega$ and so interband transition is forbidden due to Pauli blocking. At off state $\left(qV_{off} = 2E_f\right) < \hbar\omega$, so interband transition is allowed and the modulator has very low transmission. Here $V_{on}$ and $V_{off}$ are the applied voltages between graphene layers at on and off state respectively, $E_f$ is the Fermi level, $\hbar$ is the reduced plack's constant and $\omega$ is the angular frequency of incident wave.

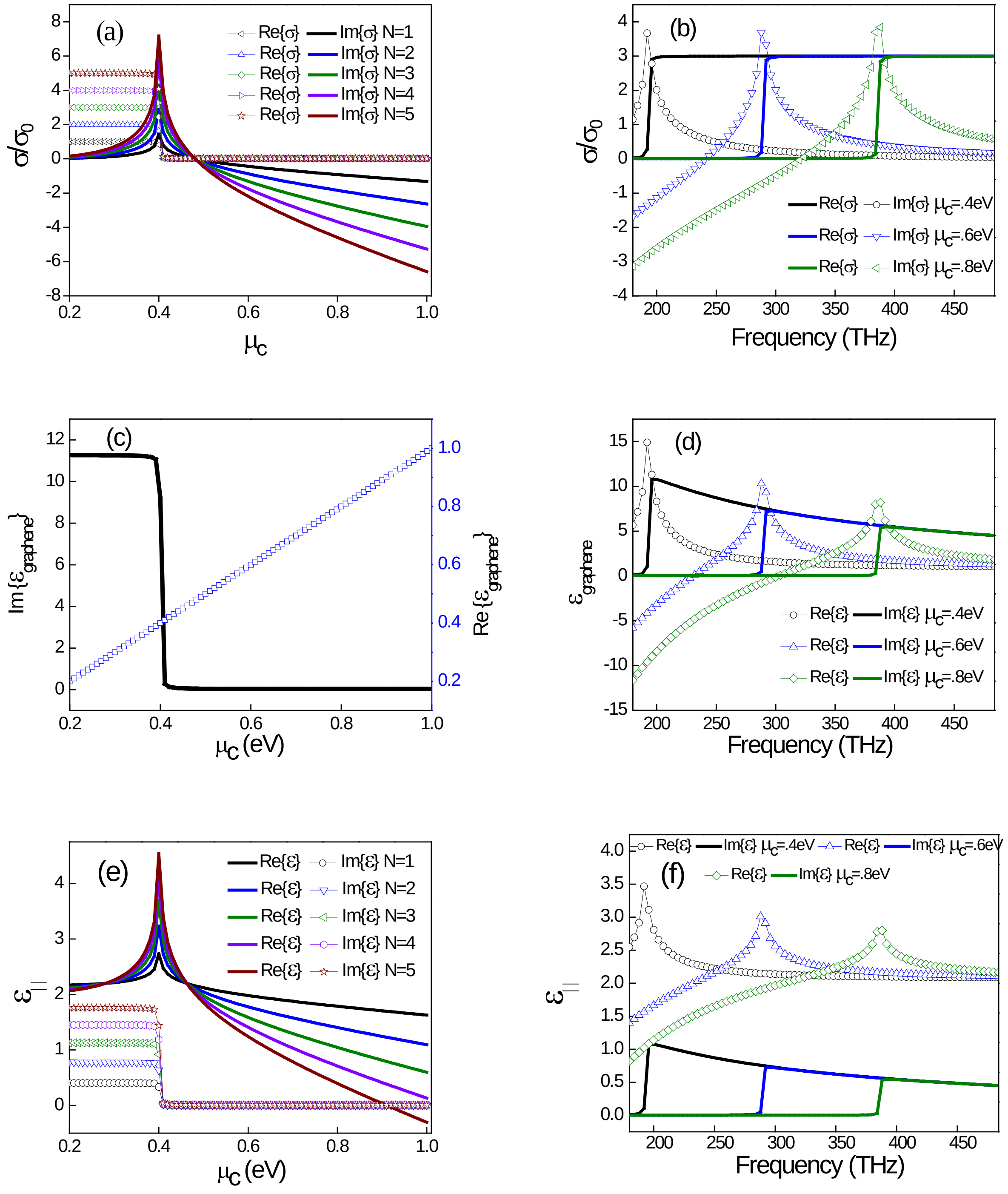

Fig. 2: (a) Normalized conductivity (real and imaginary) vs chemical potential for different number of graphene layers. (N=1 to 5) at telecommunication wavelength, $\lambda = 1.55\ \mu m$. (b) Normalized conductivity (real and imaginary) vs. frequency for different chemical potential. (c) Graphene's tangential permittivity (real and imaginary) vs. chemical potential at telecommunication wavelength, $\lambda = 1.55\ \mu m$. (d) Graphene's tangential permittivity (real and imaginary) vs. frequency for different chemical potential. (e) Parallel (tangential) permittivity (real and imaginary) of graphene-dielectric multilayer anisotropic metamaterialvs chemical potential of graphene at telecommunication wavelength, $\lambda = 1.55\ \mu m$ with N=3 (3 graphene layer in one unit cell), dielectric thickness $t_d = 10nm,$ dielectric permittivity, $\varepsilon_d = 2.2$. (f) Parallel (tangential) permittivity (real and imaginary) of graphene-dielectric multilayer anisotropic metamaterial vs. frequency for different chemical potential of graphene.

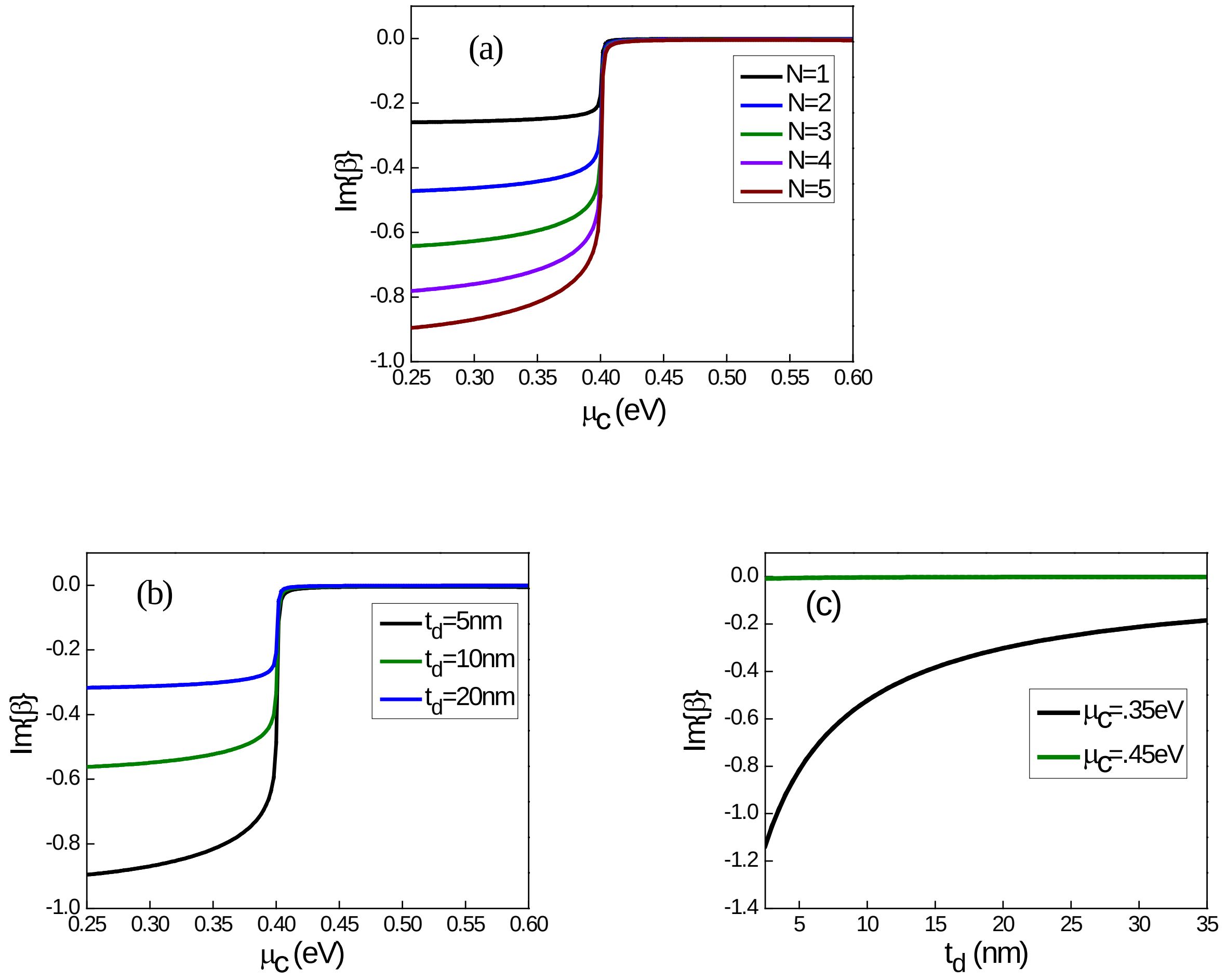

Fig. 3: Imaginary part of the propagation constant $\beta\left(k_{zp}\ in\ equation\ (14)\right)$ (a) vs. chemical potential of graphene for different number of graphene layers in each unit cell, (b) for different dielectric layer thickness. (c) vs. dielectric layer thickness for different chemical potential values of graphene.

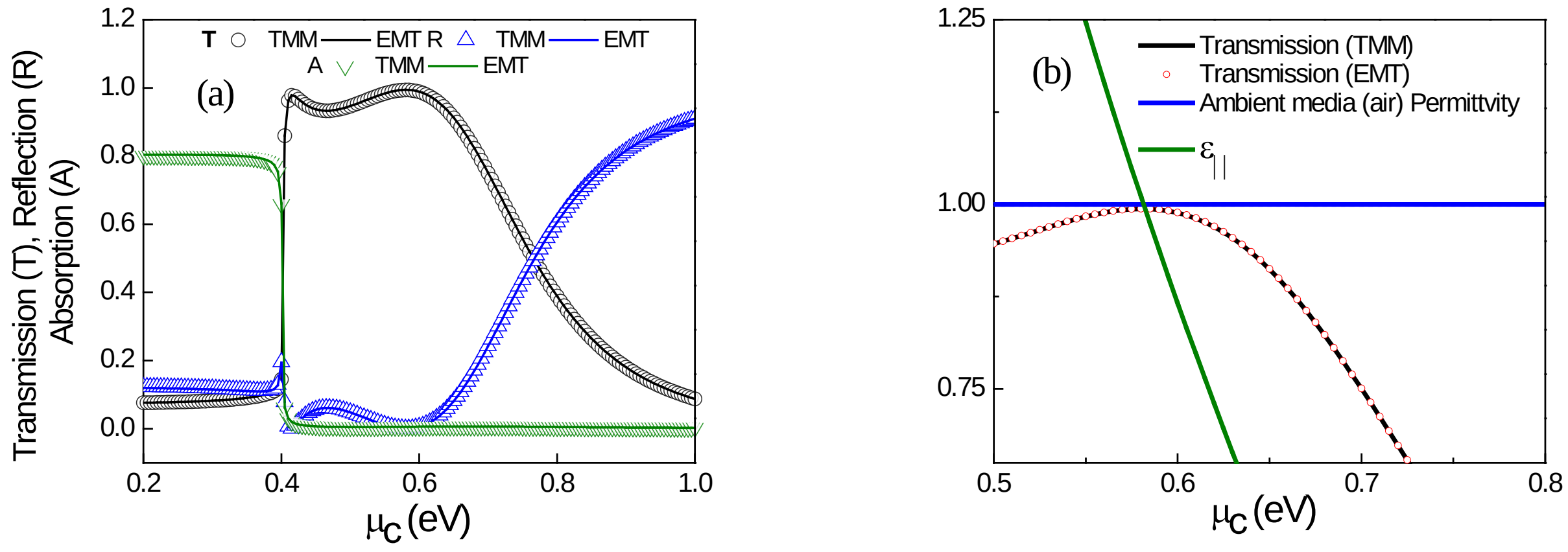

Fig. 4: (a) Transmission, reflection, absorption for (few layer) graphene-dielectric multilayer anisotropic metamaterialvs chemical potential of graphene at telecommunication wavelength $\lambda = 1.55\ \mu m$. Device length $L = 408\ nm$, number of graphene layer in a unit cell, $N = 6$, dielectric thickness $t_d = 8nm$, dielectric permittivity $\varepsilon_d = 2.2$. Symbols and solid lines represent data calculated by $4 \times 4$ transfer matrix method (equation (33a, 33b, 34b)) and by effective medium theory based analytical formulism (equation (17a, 17b, 18b)) respectively. (b) Transmission vs. chemical potential along with tangential (parallel) permittivity (equation (10)) and ambient media (air) permittivity.

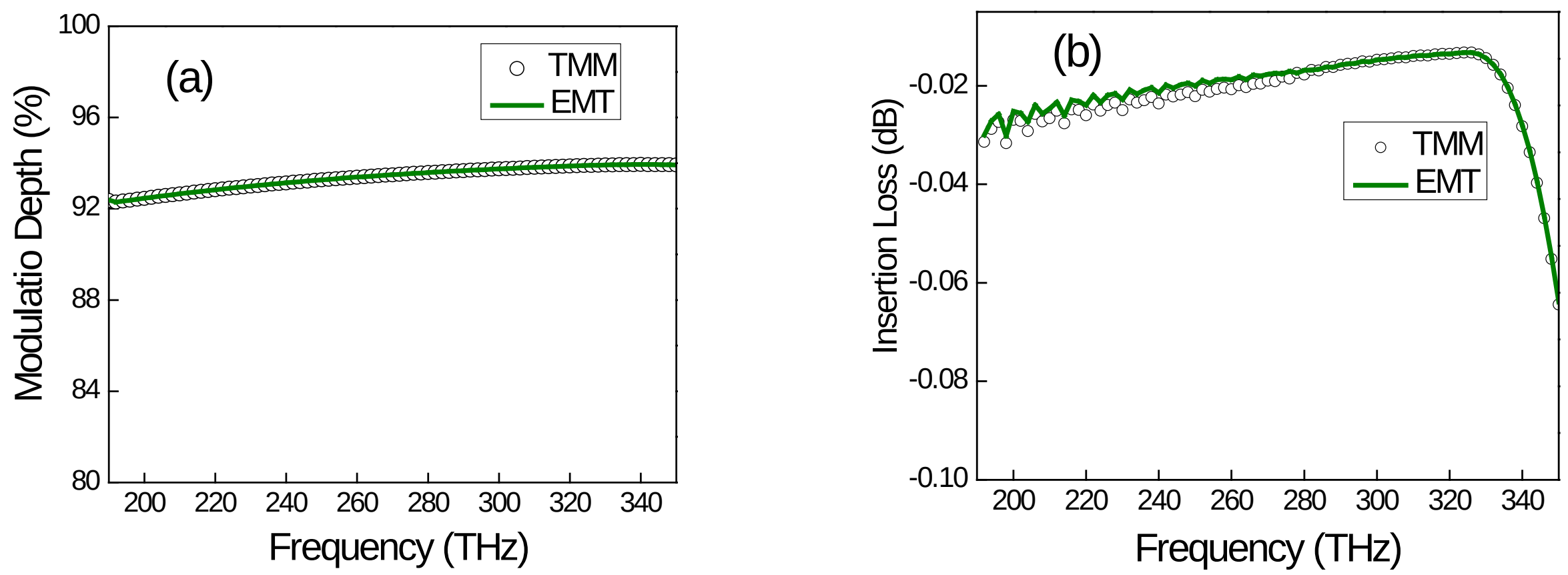

Fig. 5: (a) Modulation depth (%) (equation (19a)) for graphene-dielectric multilayer anisotropic metamaterialvs frequency. (a) Insertion loss in dB vs frequency. All device parameters are same as presented in Fig. 4. (Symbols and solid line represent data by $4 \times 4$ transfer matrix method and by effective medium theory based analytical formulism respectively).

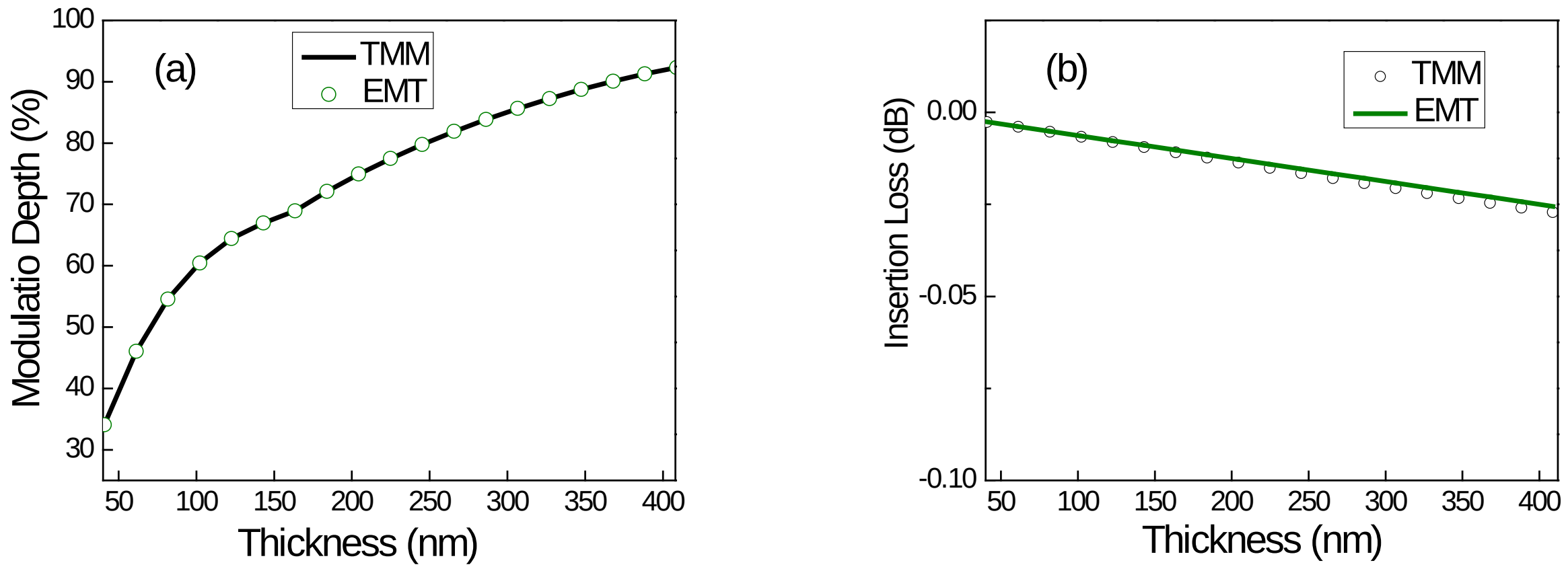

Fig. 6: (a) Modulation depth (%) (equation (19a)) for graphene-dielectric multilayer anisotropic metamaterial as a function of thickness. (a) Insertion loss in dB vs thickness. All other device parameters are same as presented in Fig. 4. (Symbols and solid line represent data by $4 \times 4$ transfer matrix method and by effective medium theory based analytical formulism respectively).